\documentclass[twocolumn,showpacs,preprintnumbers,prc]{revtex4}
\usepackage{amsmath,amssymb,amsfonts}
\usepackage{graphicx}
\newcommand{\be}{\begin{equation}}
\newcommand{\ee}{\end{equation}}
\newcommand{\bea}{\begin{eqnarray}}
\newcommand{\eea}{\end{eqnarray}}
\newcommand{\nn}{\nonumber}

\newcommand{\sls}{\!\!\!/}
\begin{document}
\title{\bf{Dead cone due to parton virtuality}}
\author{Trambak Bhattacharyya\footnote{trambakb@vecc.gov.in} and Jan-e Alam}
\medskip
\affiliation{Theoretical Physics Division, 
Variable Energy Cyclotron Centre, 1/AF, Bidhannagar, Kolkata-700064}
\date{\today}

\begin{abstract}
A general expression for the dead cone of gluons radiated by virtual partons has been derived.
The conventional dead cone for massive on-shell quarks and the dead cone for the 
massless virtual partons have been obtained by using different limits of the general expression. 
Radiative suppression due to the virtuality of initial parton jets in Heavy-Ion Collisions (HIC) 
has been discussed. It is observed that the suppression caused by the high virtuality is overwhelmingly large 
as compared to that on account of conventional dead-cone of heavy quarks. The dead cone due to virtuality
may play a crucial role in explaining the observed similar suppression patterns of light and heavy quarks
jets in heavy ion collisions at Relativistic Heavy Ion Collider (RHIC).
\end{abstract}

\maketitle

Energy loss of high energy  partons in Quark Gluon Plasma (QGP) 
which is expected to be produced in 
nuclear collisions at Large Hadron Collider (LHC) and RHIC energies 
has drawn huge attention recently. The interaction of energetic partons 
with  QGP is reflected through the measured suppression ($R_{\mathrm AA}$) of 
high transverse momentum ($p_T$) hadrons. 
The two dominant mechanisms  which cause this suppression are:
(i) elastic collision of the propagating  partons with the quarks,
anti-quarks and gluons in the thermal bath created in the HIC and 
(ii)radiation of gluons by the propagating partons due to its interactions
with the QGP. 

Classically the induced radiation takes place due to the
jiggling of the propagating particle in the medium.
Since the heavier particles jiggle less,
induced energy loss
is expected to be smaller [dead cone effect~\cite{rkellis,khar}
(see also~\cite{wang,adsw})]
for HQ compared to that for light particles.
However, the experimental data from RHIC indicates similar
energy loss by heavy quarks~\cite{stare,phenixe} 
and light partons~\cite{starl,phenixl}
in the measured kinematic range.  Various reasons  like
the anomalous mass dependence of the radiative process
due to the finite size of the QGP ~\cite{zakharov},
development of dead cone due to high virtuality of
the partons resulting from the dismantling of colour
fields during the initial hard collisions~\cite{kop}
have been proposed as the possible reasons for this observation.
The authors in~\cite{roy} concluded that
the reduction in the energy loss of HQ due to radiative
process is attributed to the dead cone effect. It is fair
to state that at this moment the issue of the observed similar suppression
of heavy and light hadrons is yet to be settled.
However, the comparable magnitude of suppression for light and heavy 
flavoured hadrons indicates that a common mechanism may be  at 
work for the radiative processes.

The dead cone is a conical domain around the direction
of motion of the massive partons within which the radiation of gluon is prohibited. 
The mass of the quark decides the size of the dead cone. However, the high energy quarks and gluons
produced from the hard collisions of the partons from the colliding nucleons 
are off-shell and their colour fields are stripped off, {\it i.e.} they have no field
to radiate. Therefore, the
partons' virtuality creates its own dead-cone which may be large depending on
the magnitude of the virtuality. The fobiden zone around the direction of motion of the 
partons due to its virtuality will be called virtual dead cone here. The conventional
dead cone (due to the mass of the quark) becomes important when the
virtuality of the quarks reduces to zero.   
The objective of the present work is to propose a quantitative estimate 
of the dead cone due to virtuality of quark jets and to investigate its interplay with the  
conventional dead cone~\cite{khar} (see also~\cite{dcone}).

The dead cone suppression obtained in~\cite{khar} actually has an analogy with radiated power distribution 
of a non-relativistic, accelerating charge particle. The average power radiated per unit solid angle is 
given by~\cite{jackson}:
\be
\frac{dP}{d\Omega}\propto |\dot{\vec{\beta}}|^2 sin^2 \theta,
\label{power}
\ee
where $\theta$ is the angle between acceleration $\dot{\vec{\beta}}$ of the particle and the direction of 
propagation of radiation, $\vec{n}$. This is a simple $sin^2 \theta$ behavior showing no radiation 
at $\theta=0~(\rm{or~n}\pi, n\in Z)$. It will be shown later that the behavior of Eq.~\ref{power}
is, indeed, similar to what one  gets for conventional dead cone~\cite{khar}.

We can take up the process $e^+ e^- \longrightarrow Q \bar{Q} g$,
(where Q denotes heavy quark) to show that the soft part of the radiative amplitude, 
when squared, gives rise to the dead-cone suppression factor of heavy quarks.
We will follow the same line of arguments as that of Ref.~\cite{kmpaul} for
virtual heavy quarks to give a formula for the suppression factor due to virtuality of quark jets. 
It will be argued that for non-zero virtuality the suppression factor is independent of the
current mass of quark. The absence of virtuality in heavy quarks results in a behavior 
similar to that obtained in~\cite{khar}; and last but not the least, we will also verify  
the absence of `conventional' dead-cone suppression for on-shell light quarks from the same formula.

\begin{figure}[h]                                                            
\begin{center}                                                               
\includegraphics[scale=0.60]{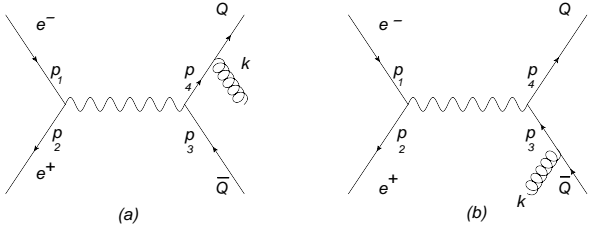}                                 
\end{center}                                                                 
\caption{Tree-level Feynman diagram for gluon radiation by (a) quark and (b) anti-quark.}
\label {fig1}                                                               
\end{figure}

The tree-level Feynman diagrams for gluon radiation by heavy quarks/anti-quarks in
the process $e^+e^-\longrightarrow Q\bar{Q}g$ are depicted in ~Fig.~\ref{fig1}.

The amplitude  for the process shown in Fig.~\ref{fig1}a can be written as:
\bea
\mathcal{M}_a&=&\frac{g_s e_Q e^2}{s} (t^a)^i_j [\bar{v}(p_2)\gamma_{\mu} u(p_1)]\nn\\
&&\times [\bar{u_i}(p_4){\epsilon \sls}^{a*}(k) \frac{(p\sls_4+k\sls +m_Q)}{((p_4+k)^2-m_Q^2)}
\gamma^{\mu}v^j(p_3)]
\label{diaa}
\eea
where $e_Q$ is the quark electric charge, e is electric charge,  $a, i, j$ are 
the color indices and $s=(p_1+p_2)^2$. 
We assume that $p_3^2=p_4^2\neq m_Q^2$. For off-shell particles
the virtuality can be defined through the parameter $V$ as:
\be
V^2=q^2-m_Q^2,
\ee  
where $q^2$ is four-momentum square of external virtual particles, $q^2=m_Q^2$ implies $V=0$, i.e.
the particle becomes on-shell. We build our calculation on very small value of V so that
the of-shell external quarks are at the vicinity of being on-shell. Under such approximations  we can use 
Dirac's equation for quark and anti-quark. 

The denominator of Eq.~\ref{diaa} can be expanded in terms of $V$ to get,

\bea
(p_4+k)^2-m_Q^2&=&p_4^2+2p_4.k+k^2-m_Q^2\nn\\
&=&V^2+2p_4.k
\eea  
with $k^2=0$ for on-shell radiated gluon.  
The numerator of Eq.~\ref{diaa} can be simplified by using the anti-commutator,
$\lbrace p_i\sls,p_j\sls\rbrace = 2p_i.p_j$
and Dirac equation $\bar{u}(p) (p\sls-m)=0$ as follows:

\bea
\bar{u_i}(p_4){\epsilon \sls}^{a^{*}}(k) (p\sls_4+k\sls +m_Q)&=&\bar{u_i}(p_4){\epsilon \sls}^{a^{*}}(k) (p\sls_4 +m_Q)\nn\\
&&+\bar{u_i}(p_4){\epsilon \sls}^{a^{*}}(k) k\sls \nn\\
&=&\bar{u_i}(p_4) (-p\sls_4+m_Q){\epsilon \sls}^{a^{*}}(k)\nn\\
&&+2p_4.{\epsilon}^{a^{*}}(k)\bar{u_i}(p_4)\nn\\
&&+\bar{u_i}(p_4){\epsilon \sls}^{a^{*}}(k) k\sls\nn\\
&=&2p_4.{\epsilon}^{a*}(k)\bar{u_i}(p_4)\nn\\
&&+\bar{u_i}(p_4){\epsilon \sls}^{a^{*}}(k) k\sls
\label{soft}
\eea
The first term of Eq.~\ref{soft} which is proportional to  $p_4.{\epsilon}^{a^{*}}(k)$ can
be defined as the soft part of amplitude~\cite{kmpaul}. For dominance of the soft 
part of $\mathcal{M}_a$ we need, $|\vec{p_i}| sin\theta>> \omega$,
where $\theta$ is the angle between parent (off-shell) quark and the gluon emitted with energy, $\omega$.
Using similar arguments for Fig.~\ref{fig1} (b), we get the total soft
amplitude given by:

\bea
[{\mathcal{M}}_{soft}]_{a+b} \sim  2\left(\frac{p_4.{\epsilon}^{a^{*}}(k)}{V^2+2p_4.k}
-\frac{p_3.{\epsilon}^{a^{*}}(k)}{V^2+2p_3.k}\right)
\label{radiation}
\eea
This part of the amplitude originates from the gluon radiation. We call it the `radiation factor', $R$. 
To obtain the cross-section of the process we square the amplitude and sum over 
the relevant colour and spin degrees of freedom.
Algebraic manipulation of $R$  gives rise to the conventional dead-cone suppression of on-shell massive quarks.
Now, with our assumption of low virtuality of heavy quarks,
R is, again, expected to lead us to a quantitative understanding of the suppression
due to off-shellness. Its interplay with the suppression due to quark mass for on-shell
quarks can be obtained simultaneously, too. 

The radiation factor R, when squared and summed over spin, gives

\bea
\sum_{\rm{spin}} |R|^2&=&\left(\frac{p_4^{\alpha}}{V^2+2p_4.k}
-\frac{p_3^{\alpha}}{V^2+2p_3.k}\right)\nn\\
&&\left(\frac{p_4^{\beta}}{V^2+2p_4.k}
-\frac{p_3^{\beta}}{V^2+2p_3.k}\right)\nn\\
&&\times\sum{\epsilon}^{a^{*}}_{\alpha}(k){\epsilon}^{a^{\prime}}_{\beta}(k)\nn\\
&=&\left(\frac{p_4^{\alpha}}{V^2+2p_4.k}
-\frac{p_3^{\alpha}}{V^2+2p_3.k}\right)\nn\\
&&\left(\frac{p_4^{\beta}}{V^2+2p_4.k}
-\frac{p_3^{\beta}}{V^2+2p_3.k}\right)\nn\\
&&\times (-\eta_{\alpha \beta}(k) \delta^{a a^{\prime}})\nn\\
&=&-\left|\frac{p_4}{V^2+2p_4.k}
-\frac{p_3}{V^2+2p_3.k}\right|^2\nn\\
&=&\frac{2p_4.p_3}{(V^2+2p_4.k)(V^2+2p_3.k)}\nn\\
&&-\frac{p_4^2}{(V^2+2p_4.k)^2}\nn\\ 
&&-\frac{p_3^2}{(V^2+2p_3.k)^2}\nn\\
&=&2R_{43}-R_{44}-R_{33}
\eea
Our problem now boils down to simplify the quantity, $2R_{43}-R_{44}-R_{33}$ to get
the necessary suppression factor.

To proceed further we take the following form of the four-momenta for the radiating and the
radiated particles:
\bea
p_4&=&E_4(1,\vec{\beta_4})\nn\\
p_3&=&E_3(1,\vec{\beta_3})\nn\\
k&=&\omega(1,\vec{n})
\eea
where $\beta_i$ is the velocity of the quark carrying a momentum 
$p_i$ and energy $E_i$ and the gluon is emitted along direction $\vec{n}$ with energy $\omega$.
If the radiating quark and antiquark were on-shell then one could write, for quark, say,

\bea
&&p_4^2=m_Q^2=E_4^2(1-\beta_4^2)\nn\\
&&\Rightarrow(1-\beta_4^2)=\frac{m_Q^2}{E_4^2}=\frac{1}{\gamma_4^2}
\eea
where $\gamma$ is the Lorentz factor. But, for our present purpose 
we define,

\bea
(1-\beta_4^2)=\frac{q^2}{E_4^2}=\frac{1}{\gamma_4^2}
\eea
which is almost but not exactly the  Lorentz factor as we have considered
off-shellness of the partons here. The factor $\omega^2 R_{43}$ can be written as:

\bea
\omega^2 R_{43}=~~~~~~~~~~~~~~~~~~~~~~~~~~~~~~~~~~~~~~~~~~~~~~~~~~~~~~~\nn\\
\frac{\omega^2 E_3 E_4(1-\beta_3\beta_4 cos\theta_{34})}
{(V^2+2E_4\omega(1-\beta_4 cos \theta_4))(V^2+2E_3\omega(1-\beta_3 cos \theta_3))}\nn\\
\eea
In the equal-$\beta$ frame, where both particles have equal velocities back to back
such that $\beta_3=\beta_4; \theta\equiv\theta_3=\theta_4-\pi; ~\rm{and} ~\theta_{34}=\theta_4-\theta_3=\pi$, 
we have $E_3=E_4=E$. In this frame we can write,

\bea
\omega^2 R_{43}=~~~~~~~~~~~~~~~~~~~~~~~~~~~~~~~~~~~~~~~~~~~~~~~~~~~~~~~\nn\\
\frac{(1+\beta^2)}
{(\frac{V^2}{\omega E}+2(1-\beta cos \theta))(\frac{V^2}{\omega E}+2(1+\beta cos \theta))}\nn\\
\label{eq12}
\eea
Similarly,

\bea
\omega^2 R_{33}=
\frac{(1-\beta^2)}
{(\frac{V^2}{\omega E}+2(1+\beta cos \theta))^2}\nn\\
\label{eq13}
\eea
and

\bea
\omega^2 R_{44}=
\frac{(1-\beta^2)}
{(\frac{V^2}{\omega E}+2(1-\beta cos \theta))^2}\nn\\
\label{eq14}
\eea
Combining Eqs.~\ref{eq12},\,\ref{eq13} and \ref{eq14} we get the
spectrum:

\bea
F&=&\omega^2(2R_{43}-R_{44}-R_{33})\nn\\
&&=4\beta^2 \left(\frac{\frac{V^4}{\omega^2 E^2}+\frac{4V^2}{\omega E}+4sin^2 \theta}
{(\frac{V^4}{\omega^2 E^2}+\frac{4V^2}{\omega E}+4(1-\beta^2 cos^2 \theta))^2}\right)
\label{deadvirt}
\eea

Next we explore different limits of $F$ given in Eq.~\ref{deadvirt}.
(i) For zero virtuality ($V=0$) of the massive quark, Eq.~\ref{deadvirt} reduces 
to the conventional dead cone factor: 
\bea
F=\omega^2(2R_{43}-R_{44}-R_{33})\longrightarrow
\frac{\beta^2 sin^2 \theta}
{(1-\beta^2 cos^2 \theta)^2}
\label{kharvirt}
\eea
This is the well-known conventional dead cone for a gluon emitted by a 
massive quark. The divergence of the factor is shielded by the quark mass
through $\beta (<1)$. 

(ii)Now we investigate the light quark limit ($\beta=1$). 
For $V=0,~\beta=1$, 
\be
F\sim\frac{1}{sin^2 \theta}
\label{light}
\ee
For light quarks Eq.~\ref{light} ensures the absence of dead-cone suppression at $\theta=0,\pi$ when
virtuality is small.

\begin{figure}[h]      
\includegraphics[scale=0.4]{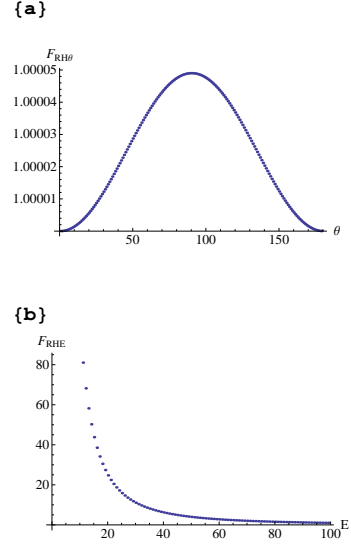}                                 
\caption{(Colour online) The variation of (a)$F_{\mathrm RH\theta}(E,\theta)$ with $\theta$ for $E=1$ GeV,
(b)$F_{\mathrm RHE}(E,\theta)$ with $E$ for $\theta=45\deg$ for heavy quarks ($\beta < 1$). 
}
\label{fig2}                                                               
\end{figure}                                                                 
For quantitative variation of $F$ with energy $E$ of the radiating partons,
and angle $\theta$ between the radiating and radiated partons,  we 
replace virtuality of heavy quark by its energy E. The emitted gluon 
carry a fraction of parent parton energy.  

In Fig.~\ref{fig2} we display the dead cone factor, $F$ for heavy quarks.
The variation of $F_{\mathrm RH\theta}=F(E=1 GeV,\theta)/F(E=1 GeV,\theta=0)$ with $\theta$ 
is depicted in Fig.~\ref{fig2}(a) for  heavy quarks ($\beta<1$). It is interesting to note that
for small $E$ and consequently for small virtuality the suppression is similar to that of a 
conventional dead cone for massive quarks.  
In Fig.~\ref{fig2} (b) the variation of $F_{\mathrm RHE}=F(E,\theta=\pi/4)/F(E=100 GeV,\theta=\pi/4)$ with $E$ 
is displayed for heavy quarks. 
 
\begin{figure}[h]      
\includegraphics[scale=0.5]{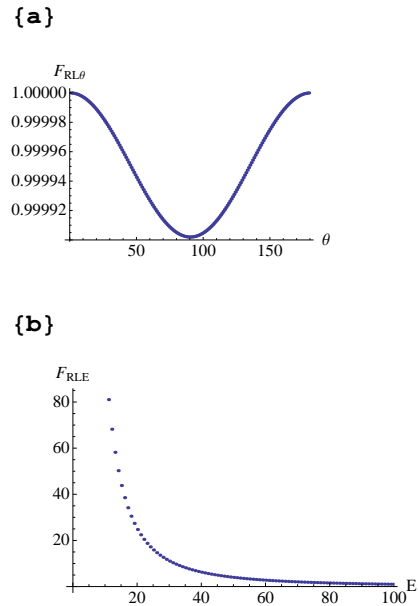}                                 
\caption{(Colour online) The variation of (a)$F_{\mathrm RL\theta}(E,\theta)$ with $\theta$ for $E=1$ GeV,
(b)$F_{\mathrm RLE}(E,\theta)$ with $E$ for $\theta=\pi/4$ for light partons ($\beta = 1$). 
}
\label{fig3}                                                               
\end{figure}

In Fig.~\ref{fig3} the dead cone factor for light quarks is illustrated. 
In Fig.~\ref{fig3} (a) the variation of $F_{\mathrm RL\theta}=F(E=1 GeV,\theta)/F(E=1 GeV,\theta=0)$ with $\theta$ 
is shown for  light partons ($\beta=1$). It is important to note that the variation of 
$F_{\mathrm RL\theta}$ with $\theta$ for light quark with low virtuality is drastically different from 
$F_{\mathrm RH\theta}$ for heavy quark.   This is obvious because for low virtuality the light 
partons are not subjected to any dead cone suppression at $\theta=0$ and $\pi$ 
unlike heavy quarks. Moreover, the $sin^{-2}\theta$ behaviour for light quarks (Eq.~\ref{light})
ensures a minimum at $\theta=\pi/2$ as opposed to a maximum at the same $\theta$ for heavy quarks. 
In Fig.~\ref{fig3} (b) the variation of $F_{\mathrm RLE}=F(E,\theta=\pi/4)/F(E=100 GeV,\theta=\pi/4)$ with $E$ 
is depicted. 
We note that the suppression is large for high $E$ (virtuality), similar to heavy quarks, 
which indicates that the dead cone effects because of large virtuality overwhelms the dead cone
due to the mass of the quarks.  


\begin{figure}[h]      
\includegraphics[scale=0.5]{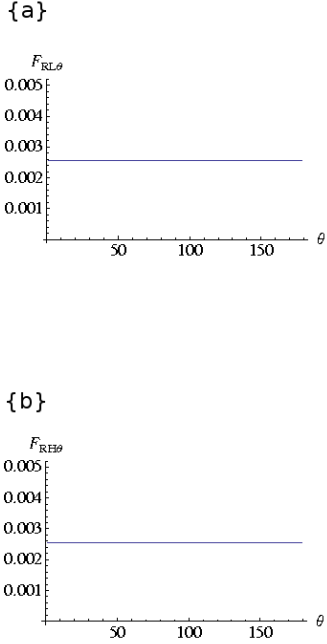}                                 
\caption{The variation of normalized $F$  with  $\theta$ for $E=20$ GeV, (a) light partons, 
(b)massive quarks. The $F$ is normalized by its value at $E=20$ GeV and $\theta=0$.
}
\label{fig4}                                                               
\end{figure}

In Figs.~\ref{fig4} (a) and (b) we plot $F_{\mathrm RL\theta}$ and $F_{\mathrm RH\theta}$
as a function of $\theta$ for $E=20$ GeV. 
We observe that the radiative suppression due to virtuality is of the same 
magnitude for both the light (Fig.~\ref{fig4} a) and heavy quarks (Fig.~\ref{fig4} b).
This leads to the conclusion that suppression due to virtuality exists irrespective of 
the current quark mass {\it i.e.} off-shell quarks, heavy or light are always exposed to 
similar radiative suppression.
It is apparent from the results displayed in Figs.~\ref{fig4}(a) and ~\ref{fig4}(b) that 
the suppression factor given by Eq~\ref{deadvirt}  may be responsible for 
the observed similar suppression patterns for light~\cite{starl,phenixl} and heavy~\cite{stare,phenixe}
hadrons of RHIC data.

In summary,  we have derived an expression for the 
dead cone of gluons radiated by virtual partons.
The conventional dead cone for massive on-shell quarks and the dead cone for the 
virtual partons have been obtained as different limits of the general expression. 
We have demonstrated the interplay between dead-cone radiative suppression of 
heavy quarks and that due to virtuality of quark jets. 
For on-shell light quarks, the absence of radiative suppression at $\theta=0,\pi$ 
is ascertained too.  We observe that the radiative suppression due to virtuality of initial parton jets 
in HIC may be overwhelmingly large as compared to that due to conventional dead-cone of heavy quarks. 
Therefore, the dead cone caused by virtuality may play a crucial role in explaining the observed similar suppression patterns 
of light and heavy quark jets in heavy ion collisions at RHIC.

\noindent{\bf Acknowledgment:}
Fruitful discussions with Surasree Mazumder,
Sourav Sarkar, Santosh K.  Das and P. Tribedy are acknowledged. TB is supported by
the Department of Atomic Energy, Govt of India.

\end{document}